\documentclass[showpacs,preprintnumbers,english,twocolumn,amsmath,amssymb]{revtex4-1}
\usepackage[T1]{fontenc}
\usepackage[latin1]{inputenc}
\usepackage{graphicx}
\usepackage{amssymb}
\usepackage{verbatim}
\usepackage{epic,eepic}
\usepackage{babel}
\usepackage{lmodern}
\usepackage{color}

\begin{document}

\title{Systematic study of  symmetry energy within the SMM picture of multifragmentation}

\author{P.~Marini$^{1,}$}\email{pmarini@comp.tamu.edu}
    \altaffiliation{Present address: Grand Acc\`el\`erateur National d'Ions Lourds, 14076 Caen, France}
\author{A.~Bonasera$^{1,2}$}
\author{G.A.~Souliotis$^{1,3}$}
\author{P. Cammarata$^{1,4}$}
\author{S.~Wuenschel$^{1,4}$}
\author{R.~Tripathi$^{1,5}$}
\author{Z. Kohley$^{1,4}$}
    \altaffiliation{Present address: National Superconducting Cyclotron Laboratory, Michigan State University, East Lansing, Michigan 48824, USA.}
\author{K.~Hagel$^{1}$}
\author{L.~Heilborn$^{1,4}$}
\author{J.~Mabiala$^{1}$}
\author{L.W. May$^{1,4}$}
\author{A.B.~McIntosh$^{1}$}
\author{S.J.~Yennello$^{1,4}$}
\affiliation{
$^{1}$Cyclotron Institute, Texas A$\&$M University, College Station, TX-77843, USA\\
$^{2}$Laboratori Nazionali del Sud, INFN, via Santa Sofia, 62, 95123 Catania, Italy\\
$^{3}$Laboratory of Physical Chemistry, Department of Chemistry, National and Kapodistrian University of Athens, 15771 Athens, Greece\\
$^{4}$ Chemistry Department, Texas A$\&$M University, College Station, TX 77843, USA\\
$^{5}$Radiochemistry Division, Bhabha Atomic Research Center, Mumbai, India\\
$^{6}$ Institute for Nuclear Research, Russian Academy of Sciences, RU-117312 Moscow, Russia 
}

\date{\today}

\begin{abstract}
A systematic study on the effect of secondary decay on the symmetry energy coefficient extracted by isoscaling and the recently proposed isobaric yield ratio methods within the Statistical Multifragmentation Model is performed. The correlations between the input symmetry energy coefficients and the calculated ones from both primary and secondary fragment yields are analysed. Results for secondary fragments show that the best estimation of the input symmetry energy coefficient within SMM is obtained by the isoscaling method, using the yields of light fragments. A comparison to experimental results is also presented.
\end{abstract}


\maketitle

 Theoretical predictions \cite{ono2004} suggest that information on the symmetry energy term  of the nuclear equation of state  can be extracted from the  isotopic distributions of primary fragments produced in multifragmentation reactions.
However, quantitative information is difficult to extract as most fragments, produced in excited states \cite{hudan03,marie98}, decay to lighter stable isotopes  on a typical time scale of $\sim 10^{-20}s$ \cite{volkov78},  before being detected. These latter fragments are commonly referred to as secondary  fragments.
Previous work has  evaluated the excitation energies of  primary fragments  \cite{hudan03,marie98} and indicated that secondary decay  may distort the signatures of the symmetry energy  contained in primary fragment observables \cite{ono2007,lefevre05}.
It is therefore important to study model predictions for observables that can be calculated for both primary and secondary fragments.

A systematic study on the effect of the secondary decay as predicted by the Statistical Multifragmentation Model (SMM) \cite{bondorf95} is presented in this report, with particular emphasis on the comparison between the recently proposed isobaric yield ratio method \cite{huang2010IYR} and the well known isoscaling \cite{xu2000,tsang2001_2,botvina2002,ono03,souliotis04,lefevre05}. 
To ascertain the degree of confidence  that can be obtained in experimental results, the correlation between input values and the quantities that can be extracted from secondary fragments was established. 
A comparison of SMM predictions to experimental data measured in $^{78,\,86}$Kr+$^{58,\,64}$Ni reactions at $35\,$MeV/nucleon with the NIMROD-ISiS array \cite{wuenschelNimrod} is also presented.

Isoscaling parameters deduced from isotopic yields measured in two similar reactions with different isotopic composition are commonly used observables  \cite{tsang2001,botvina2002,shetty2003,ono2003} to access the symmetry energy in heavy-ion collisions.
The statistical interpretation of isoscaling  
 links the isoscaling parameter $\alpha$ to the symmetry energy coefficient, $C_{sym}$, of the equation of state \cite{tsang2001,chaudhuri2008,botvina2002}:
\begin{equation}\label{eq:alphaIsosc}
\alpha=\frac{4C_{sym}(\rho)}{T}\left[\left(\frac{Z}{A}\right)_{1}^{2}-\left(\frac{Z}{A}\right)_{2}^{2}\right] = 4 \frac{C_{sym}(\rho)}{T}\Delta
\end{equation}
where $T$ is the temperature of the two fragmenting sources,  $\Delta=\left[\left(\frac{Z}{A}\right)_{1}^{2}-\left(\frac{Z}{A}\right)_{2}^{2}\right]$ and the $(Z/A)_{i}$ values correspond to the proton fraction of the n-poor ($i=1$) and n-rich ($i=2$)  sources, respectively \cite{tsang2001,botvina2002}.
Other definitions of the quantity $\Delta$ have been recently suggested \cite{ono2003,rahul,marini2011}, which take into account the fragment isotopic composition rather than the source composition. However, the debate on the proper choice of $\Delta$ is still open. In this work, we will restrict ourselves to the $\Delta$ definition suggested in \cite{tsang2001,botvina2002}, which is the one first used for the statistical interpretation of the isoscaling parameters.

The isobaric yield ratio method, recently proposed in Ref. \cite{huang2010IYR}, allows one to extract  $C_{sym}$ from the yield ratio of two pairs of isobars, $A$,  produced by the same reaction systems:
\begin{equation} \label{eq:cymsIYRLandau}
\frac{C_{sym}}{T}\approx -\frac{A}{8}\left[ lnR(3,1,A)-lnR(1,-1,A)- \delta(3,1,A)\right]
\end{equation}
where $R(3,1,A)$ and $R(1,-1,A)$ are the ratio of the yields of isobars $A$, with $N-Z=3$, $1$ and $1$, $-1$, respectively. The quantity $\delta(3,1,A)$ is the difference in the mixing entropies of isobars $A$, with $N-Z=3$, $1$ and can be neglected \cite{huang2010IYR}.

In this work the SMM model \cite{bondorf95} was used to simulate the statistical fragmentation of sources  corresponding to the Kr projectile in the reaction systems $^{78}$Kr+$^{58}$Ni  and $^{86}$Kr+$^{64}$Ni \cite{wuenschel2010} that were experimentally measured. This will allow the comparison of simulated and experimental data.
The version of SMM used for the calculation is the one described in Ref. \cite{botvina2001,ogul2011}. 
The break-up density was chosen to be $\rho = \rho_{0}/6$ \cite{tsang2001_2}. The input symmetry energy coefficient $C_{sym}^{in}$ was varied between $0$ and the standard value of the symmetry energy at normal density, $25$ MeV. 
Source excitation energies (E$^{\star}$) of $3$,  $5$ and $7$ MeV/nucleon were chosen, allowing comparison to previous experimental data \cite{wuenschel2010}.
To achieve a statistical uncertainty on the yield of every analyzed fragment better than $10\%$,
$10^{6}$ events where generated for each case analyzed in this paper. 
Isotopes for which fewer than $100$ counts were obtained in the calculation were not used in the analysis.
Isotopes with $Z>17$ were not included in the analysis, since $Z=17$ corresponds to the highest fragment charge detected with a good isotopic resolution by the NIMROD apparatus \cite{wuenschelNimrod}, used in the experimental campaign.
Final fragment yields were obtained by using the standard deexcitation procedure implemented in SMM. 
All ground and nucleon-stable excited states of light fragments, as well as 
in-medium modification of fragment properties at the break-up stage were taken into account \cite{ogul2011}. The transition energy $E_{x}^{int}$ for the beginning of restoration of the properties of isolated nuclei was chosen to be $1\,$MeV/nucleon.
Such value has, however, to be considered as a lower limit \cite{ogul2011}.

The isoscaling and the isobaric yield ratio methods were applied to  the SMM-calculated primary fragment yields including all the  isotopes for each element in the calculation. The symmetry energy coefficient to temperature ratio, $C_{sym}/T$, was extracted for each $A$  from the isoscaling parameter $\alpha$ and the isobaric yield ratios according to Eq. \ref{eq:alphaIsosc} and Eq. \ref{eq:cymsIYRLandau}, respectively. To allow a comparison of the two methods, the results $C_{sym}/T$ may be plotted as a function of $A$ for the isobaric yield ratio method and versus $A=2Z$ for isoscaling. For the isobaric yield ratio method fragments produced in the $^{78}$Kr source were considered.  
Similar results were obtained from the $^{86}$Kr source.
The symmetry energy coefficient extracted from the data analysis with both methods will be referred to as $C_{sym}^{out}$, to distinguish it from the input symmetry energy coefficient, $C_{sym}^{in}$.
\begin{figure}
\centering
\includegraphics[ width=0.95\columnwidth]{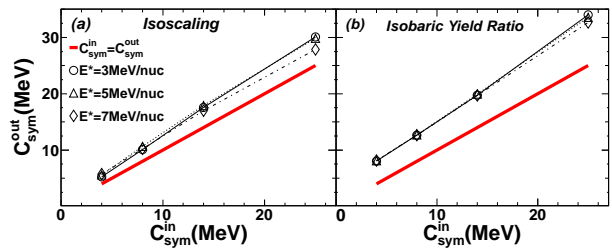} 
\caption{(Color online) Correlation between $C_{sym}^{in}$ and $C_{sym}^{out}$ values obtained from the analysis of primary fragment yields. $C_{sym}^{out}$ is obtained by the isoscaling (a) and the isobaric yield ratio (b) methods for all the $E^{\star}$ and $C_{sym}^{in}$ combinations.}
\label{fig:hot_CsymInOut}
\end{figure}
The micro-canonical temperature predicted by SMM was used as  the source temperature, $T$.
In Fig. \ref{fig:hot_CsymInOut} the weighted average of the $C_{sym}^{out}$ values obtained for each $A$ from the isoscaling (panel \textit{a}) and isobaric yield ratio method (panel \textit{b}) for different $E^{\star}$ are plotted  as a function of the input $C_{sym}^{in}$ value. 
The expected values, $C_{sym}^{out}=C_{sym}^{in}$, are also plotted for reference  (full red line).
The error bars reflect the statistical uncertainties in the fragment yield and in the temperature. 
The $C_{sym}^{out}$ values   are monotonically increasing with $C_{sym}^{in}$, independent of the method used, but the calculated values are systematically higher than the input values. The discrepancy could be ascribed to the value of the temperature used to determine $C_{sym}^{out}$. Indeed, the microcanonical temperature is different from both the ``kinetic'' \cite{DuraneTamein} and the ``chemical'' \cite{albergo85} temperatures that can be extracted from particle kinetic energies and isotopic yield ratios, respectively, and that are typically used in similar analysis \cite{lefevre05,botvina2002,zhou2011}.
 Essentially, no E$^{\star}$ dependence of $C_{sym}^{out}$ values is observed. A small deviation of the order of $10\%$ is present only for the isoscaling-extracted $C_{sym}^{out}$ at $C_{sym}^{in}=25\,$MeV for E$^{\star}=7\,$MeV/nucleon. 
 Comparing panels \textit{a} and \textit{b}, we observe that the values extracted with the two methods applied to primary fragments are similar, though a slightly larger deviation from  the expected values is obtained by the isobaric yield ratio analysis rather than the isoscaling. 
Moreover, the observed  correlations suggest that conclusions on the symmetry energy coefficient value, with a systematic uncertainty on average of $\sim20\%$ and $40\%$, could be drawn by analysing primary fragments with isoscaling and isobaric yield ratio methods, respectively.

\begin{figure}[]  
\centering
\includegraphics[width = 0.95\columnwidth, angle=0]{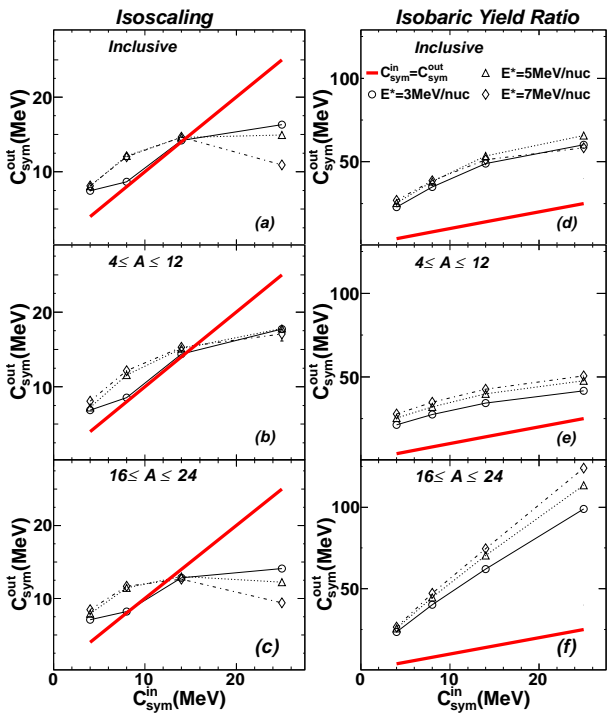} 
\caption{(Color online) Correlation between $C_{sym}^{in}$ and $C_{sym}^{out}$ values obtained from the analysis of secondary fragment yields. $C_{sym}^{out}$ is obtained by the isoscaling (a-c) and the isobaric yield ratio (d-f) methods for all the $E^{\star}$ and $C_{sym}^{in}$ combinations. In the analysis all fragment masses (a,d), $4\leq A \leq 12$ (b, e) and $16\leq A\leq 24$ (c, f) were considered. For isoscaling, $A=2Z$. }
\label{fig:CsymInOutCold}
\end{figure}

Several works aimed  to reconstruct primary fragments from measured quantities \cite{hudan03,mariniNN12}. Nevertheless, in the majority of the experiments such information is not available. Therefore we analyse the impact of the secondary decay on the $C_{sym}^{in}$ to $C_{sym}^{out}$ correlations that can be obtained with the two methods.
The same analysis procedures applied to primary fragment yields were applied to secondary fragments to build the $C_{sym}^{in}\;vs,\;C_{sym}^{out}$ correlation presented in Fig. \ref{fig:CsymInOutCold}.
Values obtained by the isoscaling and the isobaric yield ratio methods are plotted in the left and right panels, respectively. Note that the y-scale is different in the two columns. $C_{sym}^{out}$ values   were extracted  including in the calculation
 all  fragments (referred to as inclusive - panels \textit{a, d}),  
 $4\leq A \leq 12$  (panels \textit{b, e}) and   
 $16\leq A \leq 24$ (panels \textit{c, f}) fragments. We recall that $A=2Z$ for the isoscaling.
We now focus on the values obtained with the isoscaling method (panels \textit{a-c}). Largely, $C_{sym}^{out}$ increases monotonically with $C_{sym}^{in}$.
 This is true both for inclusive (panel \textit{a}) and light fragment (panel \textit{b}) isoscaling-extracted values, independent of the source $E^{\star}$. The exception is the $ E^{\star}=7\,$MeV/nucleon inclusive case, in which a $C_{sym}^{out}$ of about $10\,$MeV is obtained for $C_{sym}^{in}=25\,$MeV. This indicates that high source $E^{\star}$ might be an issue when extracting $C_{sym}^{out}$ including heavy fragments in the analysis.
 A reasonable agreement within $30\%-40\%$ of the  value of $C_{sym}^{in}$ and $C_{sym}^{out}$ is observed for $C_{sym}^{in}\leq14\,$MeV. 
Above such value, the increase of $C_{sym}^{out}$ with $C_{sym}^{in}$ is reduced for the inclusive and $4\leq A \leq 12$ cases, while a saturation is observed when applying the isoscaling to heavy fragments (panel \textit{c}). This is probably due 
to the decay toward stability of exotic isotopes. Therefore, within the SMM framework, experimental values of or close to $14\,$MeV, recently reported in \cite{shetty2005_isoscaling,lefevre05,henzlova2010,souliotis07}, should be interpreted as the lower limit for the real $C_{sym}^{in}$ value, especially if heavy fragments are included in the analysis.
 The stronger dependence between $C_{sym}^{in}$ and $C_{sym}^{out}$ when only light fragments are considered indicates that a better understanding of heavier fragment production should be sought. Reconstruction of primary fragments could provide a better understanding.
 Correlation techniques may provide a powerful tool for primary fragment reconstruction \cite{hudan03,mariniNN12}.
 A small dependence on $E^{\star}$ is observed for any mass range (except for $C_{sym}^{in}=25\,$MeV). This indicates that the different effects of secondary decay on fragments with different excitation energies mainly cancel out when taking the ratio of the same isotope produced in two similar sources

 A monotonic increase of $C_{sym}^{out}$ with $C_{sym}^{in}$ is observed for  the isobaric yield ratio method (panels \textit{d-f}). This is true independent of the mass range. A strong disagreement of a factor of up to $4-6$ is found between the input and the output values with an inclusive analysis (panel \textit{d}) and when limiting the analysis to heavy fragments (panel \textit{f}). A better agreement between $C_{sym}^{in}$ and $C_{sym}^{out}$ is obtained  for light fragments (panel \textit{e}). Although the output values are still higher than  the expected values, they present a behavior similar to the one observed for primary fragments and appear to be only shifted by a constant offset. This suggests that, with the aid of a simulation to estimate the shift, the input values could be retrieved.  In addition, a source $E^{\star}$ dependence is observed for the $C_{sym}^{out}$ values. This indicates that, when analysing experimental data, a source excitation energy selection should be applied to the data to extract meaningful information.
It is clear that, within this picture, the $C_{sym}^{out}$ values extracted with this method are strongly affected by secondary deexcitation effects and it is 
more difficult to restore the correct $C_{sym}^{in}$ for hot fragments without relying on a simulation.

\begin{figure}
\centering
\includegraphics[height=6cm,width=0.95\columnwidth]{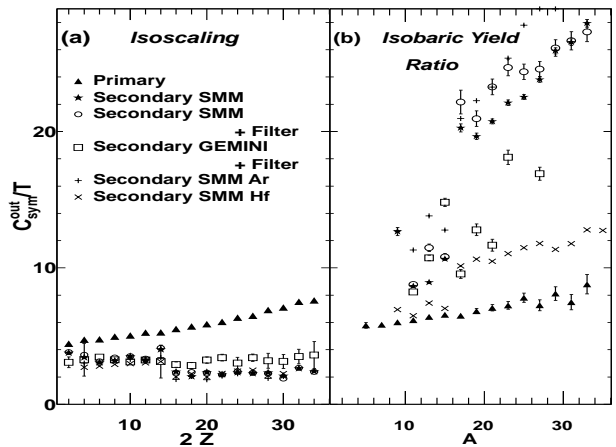} 
\caption{$C_{sym}/T$ values determined by the isoscaling (panel a) and the isobaric yield ratio method (panel b), from primary, secondary and filtered secondary yield distributions. The values obtained from GEMINI deexcitation and those obtained for  Ar and Hf sources are also plotted (see text). $C_{sym}^{in}=25\,$MeV and $E^{\star}=5\,$MeV/nucleon were used in the calculation.}
\label{fig:ColdHotSamePlot}
\end{figure}

To investigate in more detail the effect of the secondary deexcitation on the $C_{sym}^{out}$ values extracted with the two methods, we analyse the dependence of $C_{sym}^{out}/T$ on the fragment mass $A$.
In addition to results on the fragmentation of $^{78,86}$Kr sources, fragmentation of $^{39,43}$Ar and $^{156,172}$Hf sources with $E^{\star}=5\,$MeV/nucleon were also simulated by the SMM model. The four sources have the same $N/Z$ of $^{78}$Kr and $^{86}$Kr, but charges half and double the Kr charge, respectively, and were used to investigate possible finite size effects.

In Fig. \ref{fig:ColdHotSamePlot}, $C_{sym}^{out}/T$ values obtained for both primary (triangles) and secondary (stars, squares and circles) fragments  from isoscaling (panel \textit{a}) and the isobaric yield ratio (panel \textit{b}) methods are plotted  for $C_{sym}^{in}=25\,$MeV and $E^{\star}=5\,$MeV/nucleon for the Kr sources. Values of $C_{sym}^{out}/T$ obtained for secondary fragments for
$^{39,43}$Ar (crosses) and $^{156,172}$Hf (\textit{X}) sources are also plotted and will be discussed later. The secondary deexcitation  was performed, for comparison, both by SMM (stars and circles) and GEMINI (squares) \cite{charity88_2} applied to primary fragments generated by SMM.
  Indeed GEMINI   includes information on the nuclear structure and the energy levels for higher mass nuclei, in contrast to a more schematic description included in SMM \cite{bondorf95}.
The secondary yields filtered for the experimental acceptance, thresholds and resolution are labelled as ``+ Filter''  (open symbols) and are also plotted in the figure.

We now focus on the $C_{sym}^{out}/T$ values obtained for primary fragments (triangles) with both methods. As already mentioned, the values obtained with the two methods are in good agreement. 
 Moreover the $C_{sym}^{out}/T$ values show a slightly increasing trend as a function of $A$,  more pronounced for higher $C_{sym}^{in}$.
This dependence of $C_{sym}^{out}/T$ on $A$ could be attributed, according to our analysis, neither to  Coulomb contribution, nor to different average temperature at which different masses are produced.
The $C_{sym}^{out}/T$ values extracted by isoscaling from secondary fragment yields (open symbols and stars - panel a) are lower than the values extracted from primary fragment yields (triangles). Moreover  a constant behavior of $C_{sym}^{out}/T$ as a function of the fragment mass is observed, as opposed to the slightly increasing behavior observed for primary fragments. The obtained values show a gap between $A=14$ and $A=16$, not present in the value obtained by GEMINI, which is  due to the change of the deexcitation mechanism implemented in SMM for $A=16$ \cite{bondorf95}. 
The values obtained by SMM and GEMINI are in good agreement for $A<16$, within statistical uncertainty.
A good agreement of the $C_{sym}^{out}/T$ extracted from filtered (open symbols) and unfiltered (stars) secondary yield distributions with the isoscaling method is observed, which indicate that the  detection efficiency of the apparatus does not affect the values obtained with this method. 
Finally, the  $C_{sym}^{out}/T$  values obtained for the Ar and Hf sources are in good agreement with those obtained for the Kr sources. This suggests that  possible size effects, if present, also do not affect the isoscaling-extracted values.

We now focus on the values extracted with the isobaric yield ratio method (Fig. \ref{fig:ColdHotSamePlot} panel \textit{b}).
As opposed to the behavior observed for isoscaling-extracted $C_{sym}^{out}/T$ values, the $C_{sym}^{out}/T$ values determined from the yields of secondary fragments   are higher than those obtained from primary fragment yields and present an increasing trend as a function of $A$. Within the SMM model, such a trend can be ascribed to finite size effects. Indeed a much stronger increase and higher values are observed for the $C_{sym}^{out}/T$ obtained for the Ar source (crosses), while a more constant behavior and values lower than those obtained from the Kr source  are found for the Hf source (\textit{X}).  
This is in agreement with the observed finite size effects affecting the yield ratio of isobars reported in \cite{souza2012}, within canonical and grand-canonical models.
 Moreover, the values obtained by GEMINI are not in agreement with those obtained by SMM and are rather scattered, as opposed to the increasing trend observed for the SMM-calculated values, especially for $A\geq14$. Therefore,
while the effects of the secondary deexcitation are mitigated when taking the ratio of the yields of the same isotope produced by similar sources, as suggested in Ref. \cite{tsang2001_2}, this may not be the case when taking the ratio of isobars produced by the same source, as done for the isobaric yield ratio method. 
In addition, the experimental filter response causes an increase in $C_{sym}^{out}/T$  for  $13\leq A \leq 29$, as opposed to the good agreement observed for the isoscaling method. A limitation of the isobaric yield ratio method as compared to isoscaling is the  more severe influence of the background, of the relative contamination from neighbouring isotopes  for the rarest isotopes and of the identification efficiency of the experimental setup. Indeed these effects cancel out 
in the isoscaling method, while this is not true for the isobaric yield ratio method, where the ratio of the yields of different isotopes is considered.
Therefore, the isobaric yield ratio technique is more sensitive than the isoscaling
to the experimental limitations;  
the $C_{sym}^{out}$ values extracted for large masses are particularly unreliable.

\begin{figure}
\centering
\includegraphics[height=6cm,width=0.95\columnwidth]{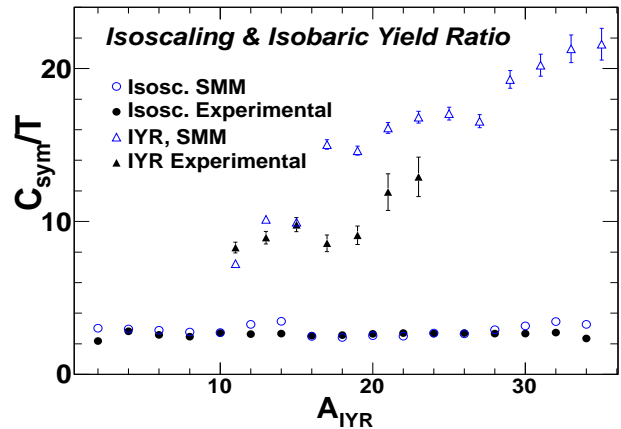} 
\caption{(Color online) Experimental (full symbols) and SMM (open symbols) $C_{sym}/T$ values obtained with isoscaling (circles) and isobaric yield ratio (triangles) methods. For the SMM calculation, $C_{sym}^{in}=14\,$MeV was used. The excitation energy was $5\,$MeV/nucleon (see text).}
\label{fig:ExpDataCfrSMM}
\end{figure}

Finally, we compare the SMM predictions to experimental data to infer, within the SMM framework, a $C_{sym}^{in}$ value.
Recently  isotopic yields of fragments produced in quasi-projectile multifragmentation of $^{78,\,86}$Kr+$^{58,\,64}$Ni at $35\,$MeV/nucleon were measured with the $4\pi$ NIMROD-ISiS array \cite{wuenschelNimrod, wuenschelNeutronBall}. Good isotopic resolution was obtained for fragments up to $Z=17$. Information on the neutron multiplicity was provided by the TAMU Neutron Ball \cite{wuenschelNeutronBall}, in which the charged particle array was housed.
The mass, charge and excitation energy of the fragmenting quasi-projectile source were reconstructed on an event-by-event basis, allowing a selection of a well defined source. The charge of the source was constrained to be $Z=30-34$.
 Details on the experimental setup and on the  source reconstruction can be found in Ref. \cite{wuenschel2010}.

It has been reported in the literature \cite{wuenschel2010,galanopoulos2010} that  better isoscaling can be obtained when selecting two sources with  well-defined  isotopic compositions rather than performing a system-to-system isoscaling.
The quasi-projectiles were therefore identified by their excitation energy per nucleon and their relative neutron excess $m_{s}=\frac{N-Z}{A}$, where $N$, $Z$ and $A$ are, respectively, the number of neutrons, protons and total nucleons in the reconstructed source.
The excitation energy of the quasi-projectile source was restricted to be between $3.5\,$MeV/nucleon and $5\,$MeV/nucleon ($\overline{E^{\star}}=4.6\,$MeV/nucleon).
Two bins in $m_{s}$ were selected, which had average values of $\overline{m_{s_{1}}}=0.097$ and $\overline{m_{s_{2}}}=0.183$ ($\overline{m_{s_{2}}} - \overline{m_{s_{1}}}=0.086$, to be compared to $0.0859$  obtained for the $^{78}$Kr  and  $^{86}$Kr sources, input of SMM).
Isoscaling was performed between the two sources, obtaining a value of 
$\Delta=0.0370$ (see Eq. \ref{eq:alphaIsosc}), consistent within $2\%$ with the $\Delta$ value used to analyze the SMM data ($\Delta=0.0378$).
The isobaric yield ratio values for $C_{sym}/T$ were extracted for a source whose average $m_{s}$ was equal to the difference in $m_{s}$ of the sources used for the isoscaling. This allows us to compare the results from the two different methods.

In Fig. \ref{fig:ExpDataCfrSMM} the experimental $C_{sym}/T$  (full symbols) and the SMM $C_{sym}^{out}/T$ prediction for secondary fragments (open symbols) obtained by the isoscaling (circles) and the isobaric yield ratio (triangles) methods are plotted as a function of the fragment mass $A$.  The values predicted by SMM were obtained for $C_{sym}^{in}=14\,$MeV and $E^{\star}=5\,$MeV/nucleon, which gives the best agreement to the experimental mass and charge distributions. We recall that a value of $C_{sym}^{in}\simeq14\,$MeV is in agreement with previous experimental results \cite{shetty2005_isoscaling,lefevre05,henzlova2010}, and should be considered as a lower limit for the real $C_{sym}^{in}$.
 The predicted (open circles) and experimental (full circles) values obtained by isoscaling method  are in good agreement, especially for masses $A>16$. We observe that SMM reproduces both the experimental trend and the values.
 On the contrary, the predicted (open triangles) and experimental (full triangles) values obtained by isobaric yield ratio method are not in agreement. This is true especially for masses $A>16$, where SMM predicts $C_{sym}^{out}/T$ values higher by a factor of about $1.5-2$ than the experimental values. Moreover  the experimental data are approximately constant as a function of $A$ for masses $A<20$. The data for mass $A=21$ and $A=23$ have a larger uncertainty due to the worsening of the isotopic resolution as the fragment mass increases. Moreover, the already mentioned sensitivity of the isobaric yield ratio method  to identification efficiency effects  makes $C_{sym}^{out}/T$ values extracted for higher masses ($A>20$) less reliable.
The increasing trend predicted by SMM is not present in the data, within our isotopic resolution.
This might suggest that a more accurate description of the deexcitation stage should be implemented in the model.

To conclude, 
 the effect of the secondary decay as predicted by SMM on the symmetry energy coefficient, $C_{sym}^{out}$, extracted by the recently proposed isobaric yield ratio method was investigated and compared to the symmetry energy coefficient obtained by isoscaling.
Monotonic increase of $C_{sym}^{out}$  with $C_{sym}^{in}$ independent of $E^{\star}$ were observed for $C_{sym}^{out}$ values extracted from primary fragment yields, suggesting that conclusions on $C_{sym}$ can be deduced from the yields of primary fragments, with both methods.  
A reasonable agreement of $C_{sym}^{in}$ and $C_{sym}^{out}$  was found for secondary fragment yields analysed with isoscaling method, in particular when limiting the analysis to lighter fragments. Experimental values of $C_{sym}\geq14\,$MeV should be considered as an upper limit, due to the saturation of the $C_{sym}^{out}$ values.
The isobaric yield ratio method retains a  strong dependence of  $C_{sym}^{out}$ on $C_{sym}^{in}$, but the obtained values are higher than the expected values by a factor of $\approx2$ in the best case.

The analysis of $C_{sym}^{out}/T$ values extracted with the two methods showed that the secondary decay affects different observables to a different degree. In particular, the isobaric yield ratio method is more vulnerable to secondary decay and detection and identification efficiency effects than the isoscaling method. This suggest that observables built from the ratio of the yields of the same fragment should be preferred to observables dependent on the absolute yields of different fragments, according to SMM predictions.
The discrepancy of SMM and GEMINI results indicates that the impact of the secondary deexcitation should be further investigated, possibly with a more refined description of the secondary decay.

Finally a comparison of SMM predictions to experimental data was presented. The SMM predictions reproduce well the $C_{sym}/T$ values obtained with the isoscaling method for $C_{sym}^{in}=14\,$MeV, in agreement with previous experimental results, which should be  considered as a lower limit for the real $C_{sym}$.

\begin{acknowledgments}
This work was supported in part by the Robert A. Welch Foundation through grant
No. A-1266, and the Department of Energy through grant No. U.S. DOE grant DE-FG03-93ER40773.
\end{acknowledgments}

\bibliography{bibliogr}

\begin{thebibliography}{10}

\bibitem{ono2004}
Akira Ono, P.~Danielewicz, W.~A. Friedman, W.~G. Lynch, and M.~B. Tsang.
\newblock {\em Phys. Rev. C}, 70:041604(R), 2004.

\bibitem{hudan03}
S.~Hudan, A.~Chbihi, J.~D. Frankland, A.~Mignon, J.~P. Wieleczko, G.~Auger,
  N.~Bellaize, B.~Borderie, A.~Botvina, R.~Bougault, et~al.
\newblock {\em Phys. Rev. C}, 67:064613, 2003.

\bibitem{marie98}
N.~Marie, A.~Chbihi, J.~B. Natowitz, A.~Le~F\`evre, S.~Salou, J.~P. Wieleczko,
  L.~Gingras, M.~Assenard, G.~Auger, Ch.~O. Bacri, et~al.
\newblock {\em Phys. Rev. C}, 58:256, 1998.

\bibitem{volkov78}
V.V. Volkov.
\newblock {\em Phys. Rep.}, 44:93, 1978.
\newblock and Ref. therein.

\bibitem{ono2007}
A.~Ono.
\newblock {\em AIP Conf. Proc.}, 884:292, 2007.

\bibitem{lefevre05}
A.~Le~F\`evre, G.~Auger, M.~L. Begemann-Blaich, N.~Bellaize, R.~Bittiger,
  F.~Bocage, B.~Borderie, R.~Bougault, B.~Bouriquet, J.~L. Charvet, et~al.
\newblock {\em Phys. Rev. Lett.}, 94:162701, 2005.

\bibitem{bondorf95}
J.P. Bondorf, A.S. Botvina, A.S. Iljinov, I.N. Mishustin, and K.~Sneppen.
\newblock {\em Phys. Rep.}, 257:133, 1995.

\bibitem{huang2010IYR}
M.~Huang, Z.~Chen, S.~Kowalski, Y.~G. Ma, R.~Wada, T.~Keutgen, K.~Hagel,
  M.~Barbui, A.~Bonasera, C.~Bottosso, et~al.
\newblock {\em Phys. Rev. C}, 81:044620, 2010.

\bibitem{xu2000}
H.~S. Xu, M.~B. Tsang, T.~X. Liu, X.~D. Liu, W.~G. Lynch, W.~P. Tan,
  A.~Vander~Molen, G.~Verde, A.~Wagner, H.~F. Xi, et~al.
\newblock {\em Phys. Rev. Lett.}, 85:716, 2000.

\bibitem{tsang2001_2}
M.~B. Tsang, C.~K. Gelbke, X.~D. Liu, W.~G. Lynch, W.~P. Tan, G.~Verde, H.~S.
  Xu, W.~A. Friedman, R.~Donangelo, S.~R. Souza, et~al.

\bibitem{botvina2002}
A.~S. Botvina, O.~V. Lozhkin, and W.~Trautmann.
\newblock {\em Phys. Rev. C}, 65:044610, 2002.

\bibitem{ono03}
Akira Ono, P.~Danielewicz, W.~A. Friedman, W.~G. Lynch, and M.~B. Tsang.
\newblock {\em Phys. Rev. C}, 68:051601(R), 2003.

\bibitem{souliotis04}
G.A. Souliotis, M.~Veselsky, D.V. Shetty, and S.J. Yennello.
\newblock {\em Phys. Lett. B}, 588:35, 2004.

\bibitem{wuenschelNimrod}
S.~Wuenschel, K.~Hagel, R.~Wada, J.B. Natowitz, S.J. Yennello, Z.~Kohley,
  C.~Bottosso, L.W. May, W.B. Smith, D.V. Shetty, et~al.
\newblock {\em Nucl. Instr. Meth. A}, 604:578, 2009.

\bibitem{tsang2001}
M.~B. Tsang, W.~A. Friedman, C.~K. Gelbke, W.~G. Lynch, G.~Verde, and H.~S. Xu.
\newblock {\em Phys. Rev. Lett.}, 86:5023, 2001.

\bibitem{shetty2003}
D.~V. Shetty, S.~J. Yennello, E.~Martin, A.~Keksis, and G.~A. Souliotis.
\newblock {\em Phys. Rev. C}, 68:021602, 2003.

\bibitem{ono2003}
Akira Ono, P.~Danielewicz, W.~A. Friedman, W.~G. Lynch, and M.~B. Tsang.
\newblock {\em Phys. Rev. C}, 68:051601(R), 2003.

\bibitem{chaudhuri2008}
G.~Chaudhuri, S.~Das Gupta, and M.~Mocko.
\newblock {\em Nucl. Phys. A}, 813:293, 2008.
\newblock and Refs. therein.

\bibitem{rahul}
R.~Tripathi, A.~Bonasera, S.~Wuenschel, L.~W. May, Z.~Kohley, G.~A. Souliotis,
  S.~Galanopoulos, K.~Hagel, D.~V. Shetty, K.~Huseman, et~al.

\bibitem{marini2011}
P.~Marini, A.~Bonasera, A.~McIntosh, R.~Tripathi, S.~Galanopoulos, K.~Hagel,
  L.~Heilborn, Z.~Kohley, L.~W. May, M.~Mehlman, S.~N. Soisson, G.~A.
  Souliotis, D.~V. Shetty, W.~B. Smith, B.~C. Stein, S.~Wuenschel, and S.~J.
  Yennello.
\newblock {\em Phys. Rev. C}, 85:034617, 2012.

\bibitem{wuenschel2010}
S.~Wuenschel, A.~Bonasera, L.W. May, G.A. Souliotis, R.~Tripathi,
  S.~Galanopoulos, Z.~Kohley, K.~Hagel, D.V. Shetty, K.~Huseman, et~al.
\newblock {\em Nucl. Phys. A}, 843:1, 2010.
\newblock S.~Wuenschel, Ph.D. Thesis, Texas A$\&$M University, 2009.

\bibitem{botvina2001}
A.~S. Botvina and I.~N.Mishustin.
\newblock {\em Phys. Rev. C}, 63:061601(R), 2001.

\bibitem{ogul2011}
R.~Ogul \textit{et al.}
\newblock {\em Phys. Rev. C}, 83:024608, 2011.

\bibitem{DuraneTamein}
D.~Durand and B.~Tamain.
\newblock \textit{Notes from course held at the International Nuclear Physics
  School Joliot-Curie} ({F}rance).

\bibitem{albergo85}
E.~Costanzo A.~Rubbino S.~Albergo, S.~Costa.
\newblock {\em Il Nuovo Cimento A}, 89:1, 1985.

\bibitem{zhou2011}
P.~Zhou, W.~D. Tian, Y.~G. Ma, X.~Z. Cai, D.~Q. Fang, and H.~W. Wang.
\newblock {\em Phys. Rev. C}, 84:037605, 2011.

\bibitem{mariniNN12}
P.~Marini, A.~Bonasera, A.~B. McIntosh, M.~Boisjoli, P.~Wigg, and A.~Chbihi.
\newblock Isotopic yields as a probe of the symmetry energy: dealing with
  secondary decay effects.
\newblock 2012.
\newblock Nucleus-Nucleus Conference Proceedings.

\bibitem{shetty2005_isoscaling}
D.~V. Shetty, A.~S. Botvina, S.~J. Yennello, G.~A. Souliotis, E.~Bell, and
  A.~Keksis.
\newblock {\em Phys. Rev. C}, 71:024602, 2005.

\bibitem{henzlova2010}
D~Henzlova, A.~S. Botvina, K.H. Schmidt, V.~Henzl, P.~Napolitani, and M.~V.
  Ricciardi.
\newblock {\em J. Phys. G: Nucl. Part. Phys.}, 37:085010, 2010.

\bibitem{souliotis07}
G.~A. Souliotis, A.~S. Botvina, D.~V. Shetty, A.~L. Keksis, M.~Jandel,
  M.~Veselsky, and S.~J. Yennello.
\newblock {\em Phys. Rev. C}, 75:011601, 2007.

\bibitem{charity88_2}
R.J. Charity, M.A. McMahan, G.J. Wozniak, R.J. McDonald, L.G. Moretto, D.G.
  Sarantites, L.G. Sobotka, G.~Guarino, A.~Pantaleo, L.~Fiore, et~al.
\newblock {\em Nucl. Phys. A}, 483:371, 1988.

\bibitem{souza2012}
S.~R. Souza and M.~B. Tsang.
\newblock {\em Phys. Rev. C}, 85:024603, 2012.

\bibitem{wuenschelNeutronBall}
S.~Wuenschel \emph{et al.}
\newblock {\em AIP Conf. Proc.}, 1099:816, 2009.

\bibitem{galanopoulos2010}
S.~Galanopoulos, G.A. Souliotis, A.L. Keksis, M.~Veselsky, Z.~Kohley, L.W. May,
  D.V. Shetty, S.N. Soisson, B.C. Stein, S.~Wuenschel, and S.J. Yennello.
\newblock {\em Nucl. Phys. A}, 837:145, 2010.

\end{thebibliography}
\bibliographystyle{unsrt}

\end{document}